\documentclass[prl,twocolumn,superscriptaddress,showpacs,preprintnumbers,amsmath,amssymb]{revtex4}
\usepackage{amssymb}

\usepackage{graphicx}
\usepackage{dcolumn}
\usepackage{bm}
\usepackage{epsfig}
\newcommand{\ignore}[1]{}

\begin{document}


\title{Tuning Fe Nucleation Density with Charge Doping of Graphene Substrate}

\author{Wenmei Ming}
\affiliation{Department of Materials Science and Engineering,
University of Utah, Salt Lake City, UT 84112}

\author{Feng Liu}
\thanks{Corresponding author. E-mail: fliu@eng.utah.edu}
\affiliation{Department of Materials Science and Engineering,
University of Utah, Salt Lake City, UT 84112}

\date{\today}


\begin{abstract}
We have demonstrated that the island nucleation in the initial stage
of epitaxial thin film growth can be tuned by substrate surface
charge doping. This charge effect was investigated using spin
density functional theory calculation in Fe-deposition on graphene
substrate as an example. It was found that hole-doping can
apparently increase both Fe-adatom diffusion barrier and Fe
inter-adatom repulsion energy occurring at intermediate separation,
whereas electron-doping can decrease Fe-adatom diffusion barrier but
only slightly modify inter-adatom repulsion energy. Further kinetic
Monte Carlo simulation showed that the nucleation island density can
be increased up to ten times larger under hole-doping and can be
decreased down to ten times smaller than that without doping. Our
findings indicates a new route to tailoring the growth morphology of
magnetic metal nanostructure for spintronics applications via
surface charge doping.
\end{abstract}

\maketitle

Because of the ideal two-dimensional honeycomb crystal structure and
exotic linear dispersed electronic band structure, graphene has
attracted intensive research effort of surface functionalization
with external adsorbates in order to incorporate carrier doping
\cite{carrier1,carrier2}, magnetism \cite{mag1,mag2}, catalysis
\cite{catalyst1,catalyst2} and superconductivity \cite{super1,
super2}, which are strongly related to the bonding involving orbital
hybridization and charge transfer between adsorbate and graphene.
Due to its only one-atomic thickness, epitaxial graphene is usually
unintentionally doped with finite concentration of free carriers
through substrate charge transfer \cite{subc1}. Wider charge doping
can also be realized via electric field effect \cite{Efie} or
substrate doping \cite{subdop}. The resulting charge effect, on one
hand may alter the bonding strength between adsorbate and graphene,
affecting Fe adsorption and diffusion \cite{cbond}; on the other
hand, it may modulate the adsorbate-adsorbate interaction
\cite{adad}, affecting adsorbate island nucleation. Similar
electronic tailing of adsorbate-substrate and adsorbate-adsorbate
interactions were observed experimentally on ultrathin oxide film
supported on metallic substrate by varying the thickness of the
oxide film \cite{oxideT1}.

For weakly corrugated metallic surfaces such as M/M(111) (M=Al, Cu,
Ag, Au)) \cite{corr1, corr2, corr3, corr4} the perturbation to the
adsorbate diffusion barrier due to the existence of surrounding
adsorbates beyond the nearest-neighbor distance is comparable to the
adsorbate diffusion barrier. The resulting inter-adsorbate repulsion
part at intermediate distance leads to effective increase of
diffusion barrier, giving rise to the significantly larger
nucleation island density observed than from mean-field nucleation
theory, which includes only nearest-neighbor interaction. Recent
experiment of Fe deposited on epitaxial graphene on 6H-SiC(0001)
\cite{Fehigh} reported that island density increased almost linearly
with depositon amount up to 2.5~ML without appearance of saturation
and showed weak temperature dependence. These are the indications of
graphene being another weakly corrugated system for Fe with sizeable
inter-adatom repulsion at distance larger than nearest-neighbor
distance. Further DFT calculation predicts the electronic origin of
the Fe-Fe repulsion \cite{FeFerepul}.

In this work we are motivated to study the charge doping effects on
the Fe adsorption, diffusion and adatom-adatom interaction on
graphene substrate. We found that hole-doping increases the
adsorption energy, diffusion barrier and Fe-Fe repulsion energy, and
that electron-doping decreases the diffusion barrier but only
modifies slightly the adsorption energy and Fe-Fe repulsion energy.
It is therefore expected that higher Fe island density can be
achieved by hole doping and more layer-like film can be achieved by
electron doping. Further kinetic Monte Carlo (kMC) simulations shows
that Fe nucleation island density can be tuned from being six times
larger under hole doping to being ten times smaller under electron
doping than the zero-doping case. This wide-range tunability may
provide the potential to grow Fe film with island morphology as
magnetic storage device and more uniform layer morphology as
magnetic electric contact for spin injection in spintronic
applications.

The spin density functional theory (DFT) calculations were performed
by using projector augment wave pseudopotential (PAW) \cite{PAW}
with the generalized gradient approximation (GGA) \cite{GGA} to the
exchange-correlation functional, as implemented in VASP package
\cite{vasp}. $7\times7$ graphene supercell plus 13~{\AA} vacuum was
used as the substrate. 400~eV energy cutoff and $3\times3\times1$
$\Gamma$-centered k-mesh were used for wavefuntion expansion and
k-space integration, respectively. Charge doping was simulated by
adding (removing) electron for electron (hole) doping and
compensating opposite charge background to keep the system neutral.
The charge was varied from hole concentration of
$-1.17\times10^{14}~e/cm^2$ to electron concentration of
$0.78\times10^{14}~e/cm^2$. One Fe adatom was used to calculate the
adsorption energy, diffusion barrier and magnetic property. Two Fe
atoms with varying separation were used to calculate the inter-atom
interaction energy. All the structures were relaxed in terms of
internal atomic coordinates using conjugate gradient method until
the force exerted on each atom is smaller than 0.01~eV$/{\AA}^3$.
The transition saddle point along adatom diffusion path was
identified using nudged elastic band method \cite{NEB}.

\begin{figure}
\includegraphics[clip,scale=0.25]{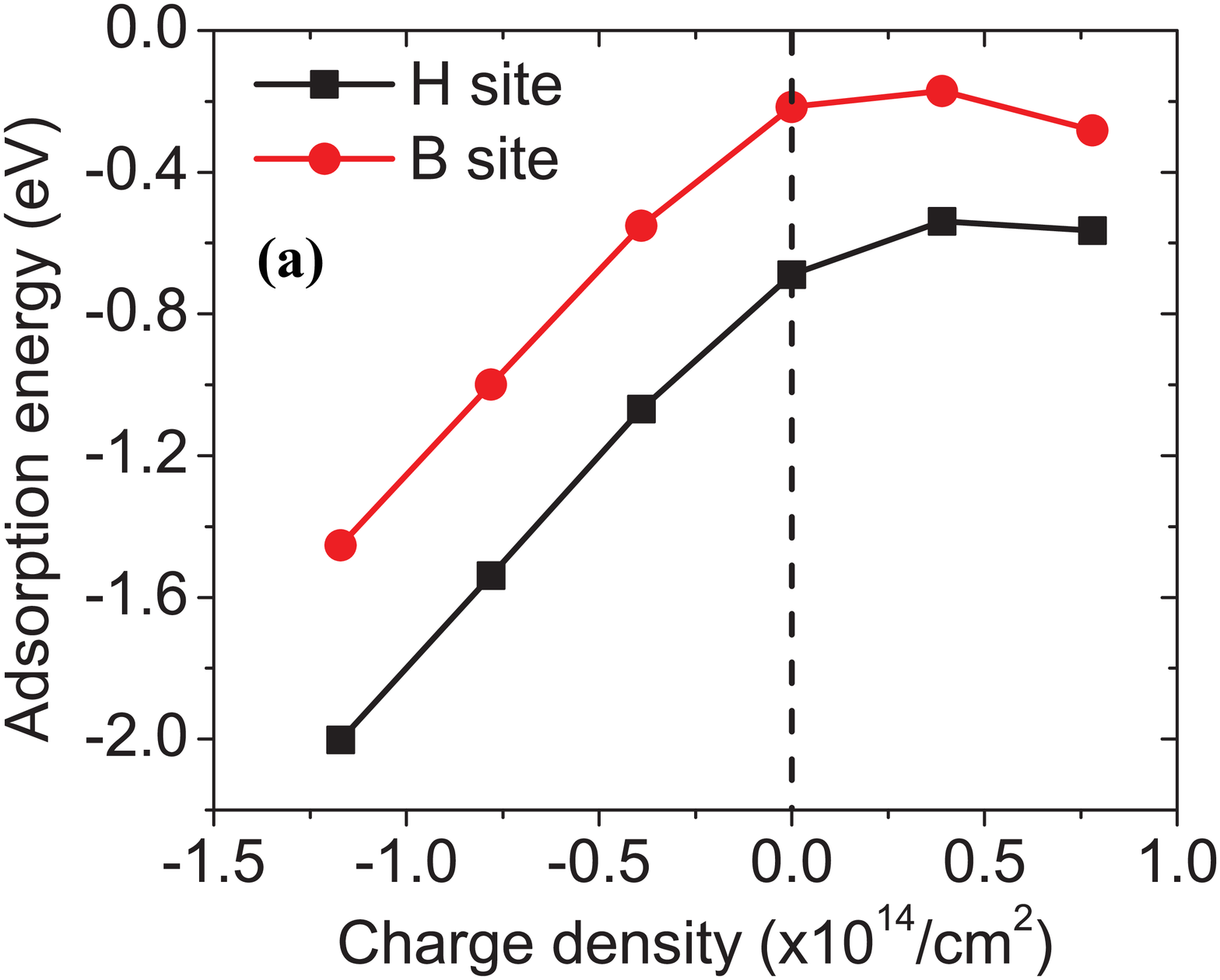}
\includegraphics[clip,scale=0.25]{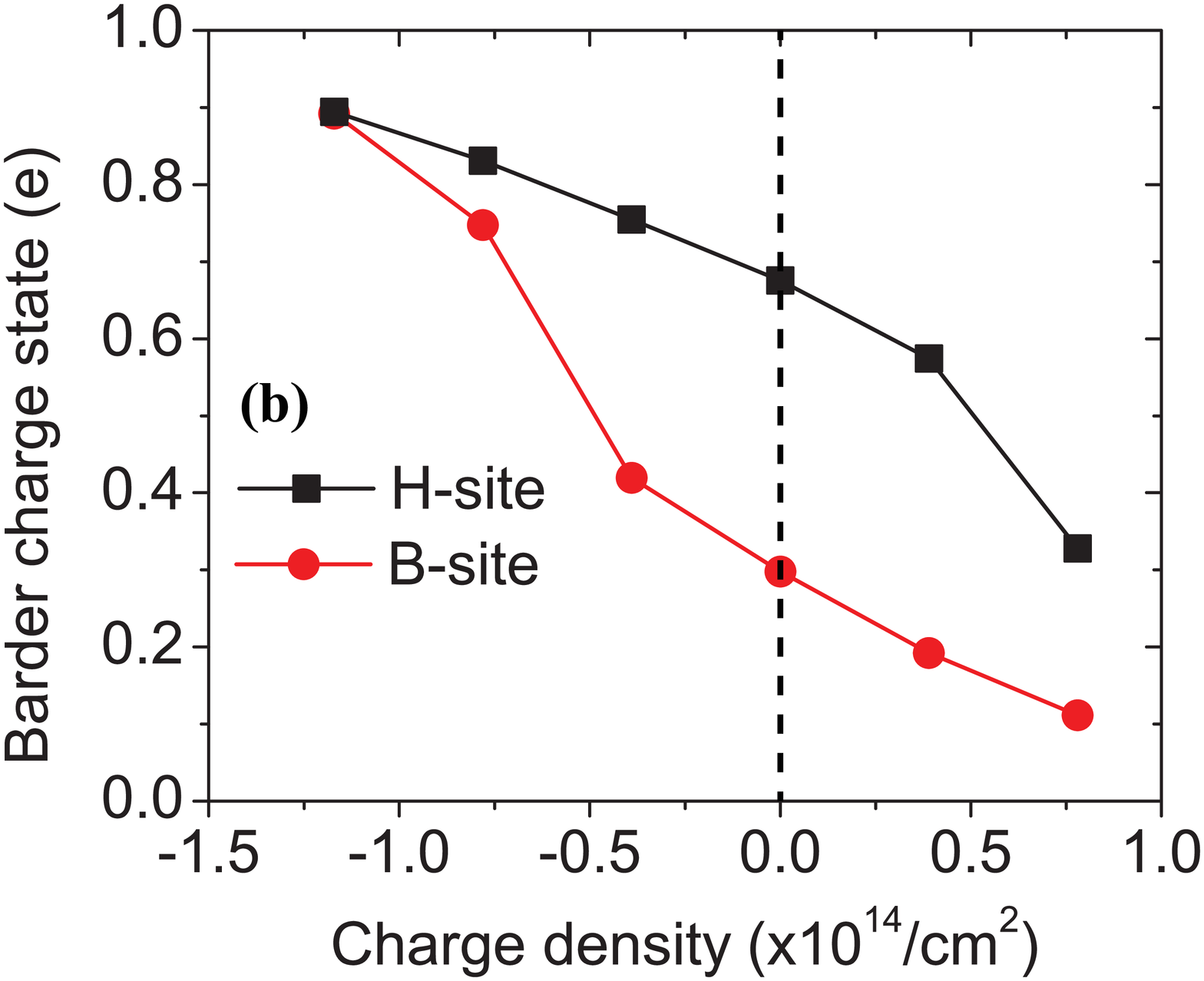}
\includegraphics[clip,scale=0.25]{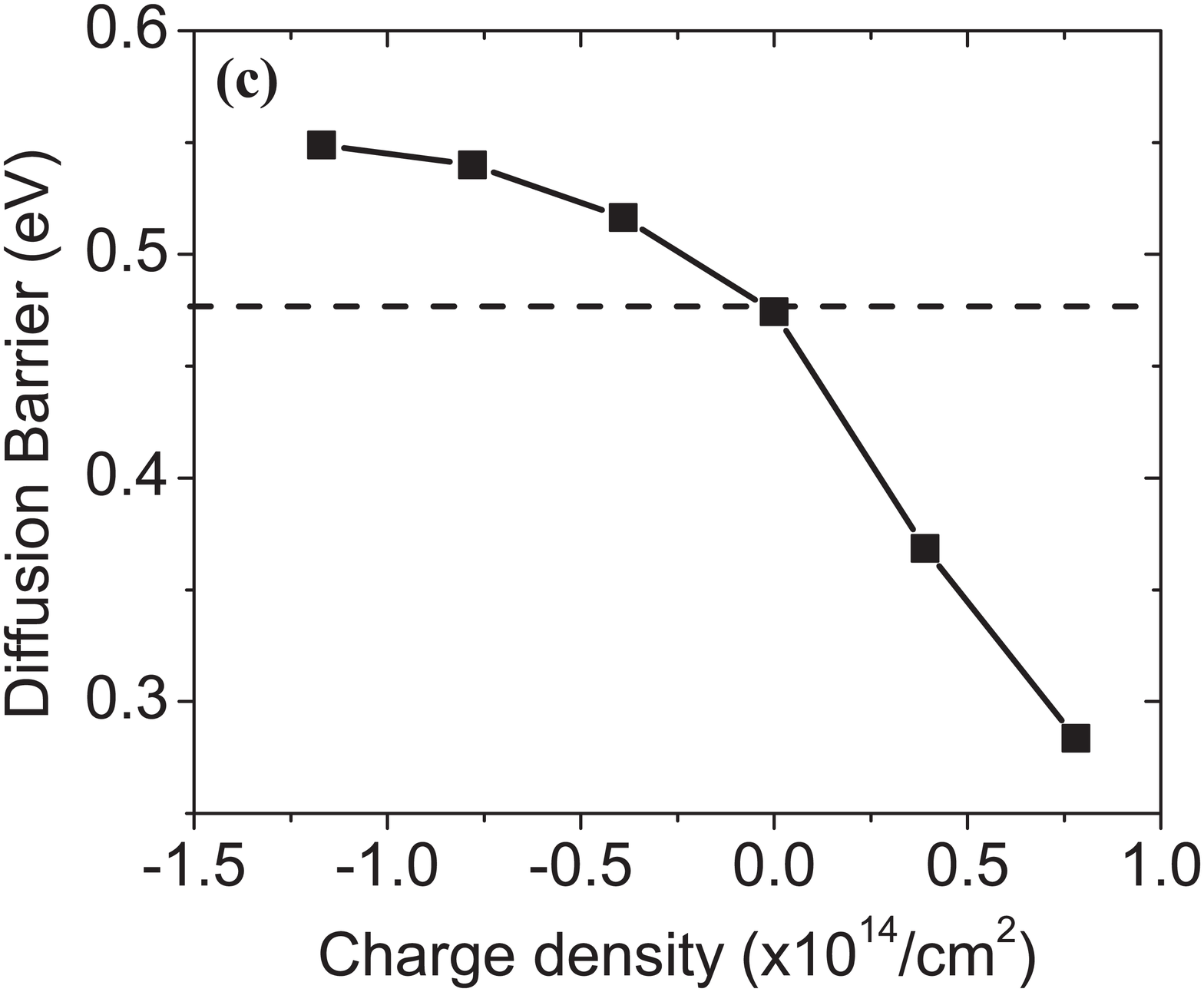}
\caption{(color online) (a) Adsorption energy versus charge doping
concentration for Fe adatom at H-site and B-site; (b) Bader charge
of Fe adatom at H-site and B-site versus charge doping
concentration; (c) diffusion barrier versus charge doping
concentration.}
\end{figure}

First we found that within the doping concentration considered here
Fe adsorption site is the hollow site (H-site) and the transition
saddle point is the bridge site (B-site). The adsorption energy
$E_{ad}$ is defined as $E_{ad}=E(Graphene+Fe)-E(Graphene)-E(Fe)$,
where $E(Graphene+Fe)$ is the energy of adatom+graphene, E(Graphene)
is the energy of clean graphene with charge doping and E(Fe) is the
energy of isolated Fe atom. It is plotted as a function of charge
doping concentration for both Fe at H-site and Fe at B-site in Fig.
1(a). With respect to zero-charge doping case hole doping increases
rapidly the adsorption energy but electron doping only slightly
changes the adsorption energy. During the process of Fe adsorption
on graphene, it has graphene $\pi$ bond breaking and Fe-C bond
formation, so the adsorption energy will be proportional to the bond
energy difference between Fe-C and graphene $\pi$. The charge doping
dependence of Fe-C bond energy and graphene $\pi$ bond energy will
give rise to the trend of Fe-adsorption energy variation as a
function of charge doping concentration. For graphene $\pi$ bond it
will have lower bond energy because less bonding states will be
occupied under hole doping, and also lower bond energy because more
anti-bonding states will be occupied under electron doping. For Fe-C
bond, it involves charge transfer and orbital hybridization. The
energy gain due to the charge transfer is proportional to the
difference between electron energy levels of Fe atom before
adsorption and the Fermi energy of graphene. For clean graphene
electron doping decreases its Fermi energy and hole doping increases
its Fermi energy, so the difference between electron energy levels
of isolated Fe atom and substrate Fermi energy will become larger
for hole doping indicating charge transfer from Fe to grapene will
be easier, but smaller for electron doping indicating charge
transfer from Fe to graphene will be blocked. Therefore the combined
effect of graphene $pi$ bond breaking and charge transfer may
increase Fe adatom adsorption energy with hole doping but only
slightly varies with electron doping.

We further calculated the change of Fe adatom charge transfer in
response to graphene work function change (equivalently Fermi energy
change) under charge doping in Fig. 1(b). The amount of charge
transfer from Fe adatom to graphene is represented by Bader charge.
As argued above, there are more charge transfer under hole doping
and less charge transfer for both Fe at H-site and B-site. We may
estimate the adsorption energy gain from charge transfer part using
the following model:
\begin{eqnarray}
E_{ad}(q)=E_r(q)-q\phi
\end{eqnarray}
where $q$ is the adatom charge transfer, $\phi$ is the graphene
substrate work function and $E_r$ is remaining contribution to the
adsorption energy. The $E_{ad}$ variation due to the change of
$\phi$ can be estimated with respect to $\phi$ of no-doping graphene
using:
\begin{eqnarray}
\Delta E_{ad}(q)=\Delta E_r(q)-q\Delta\phi-\phi\Delta q
\end{eqnarray}
Three contributions are include in the variation of $E_{ad}$. While
it is not clear to see in what fashion the first one $\Delta E_r(q)$
changes $E_{ad}$, we can easily see that the second term increases
$E_{ad}$ in hole doping when the work function increases but
decreases $E_{ad}$ in electron doping when the work function
decreases. Similarly the third term increases $E_{ad}$ in hole
doping when the charge transfer $q$ is increased but decreases
$E_{ad}$ in electron doping when the charge transfer $q$ is
decreased.

\begin{figure}
\includegraphics[clip,scale=0.25]{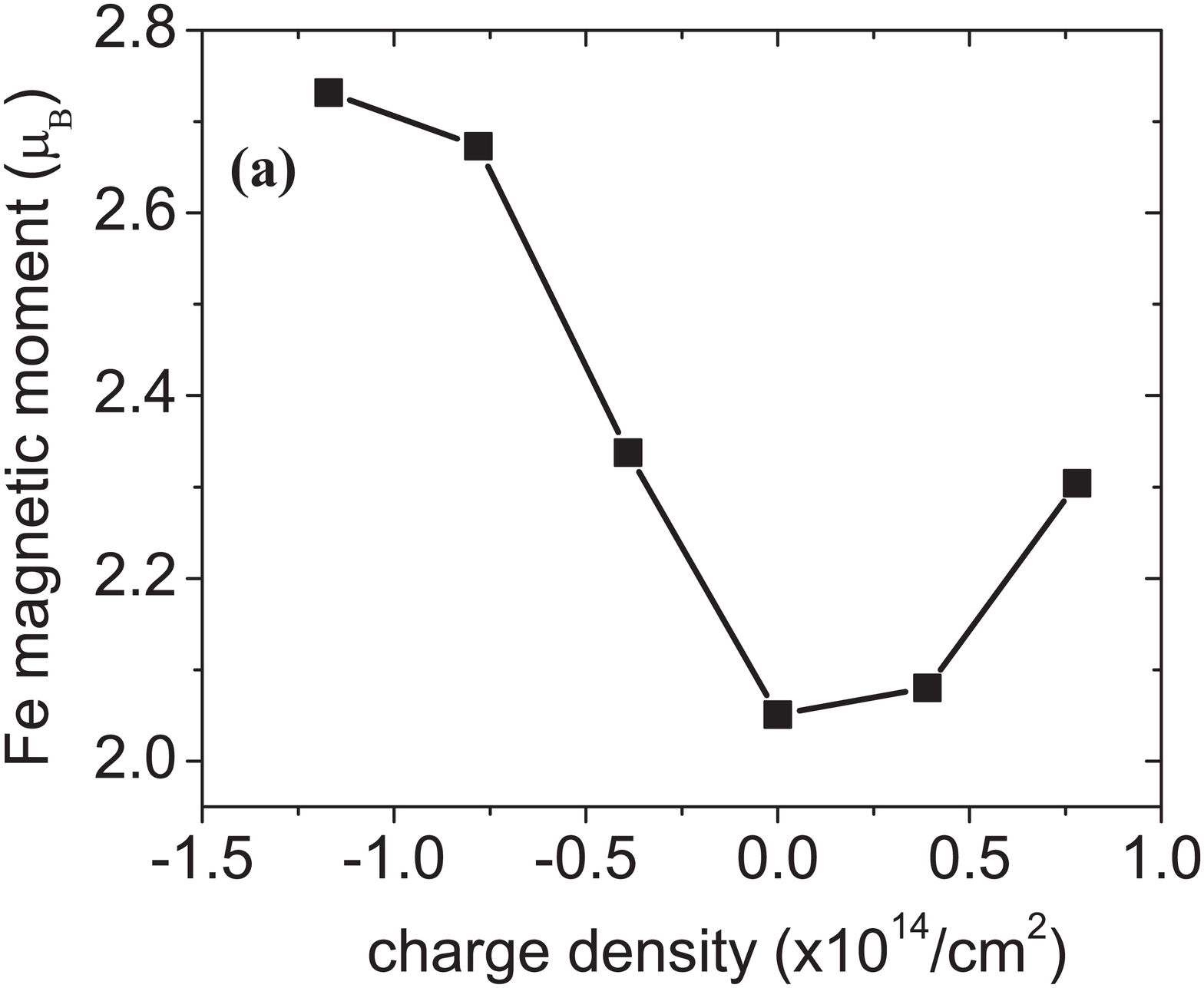}
\includegraphics[clip,scale=0.40]{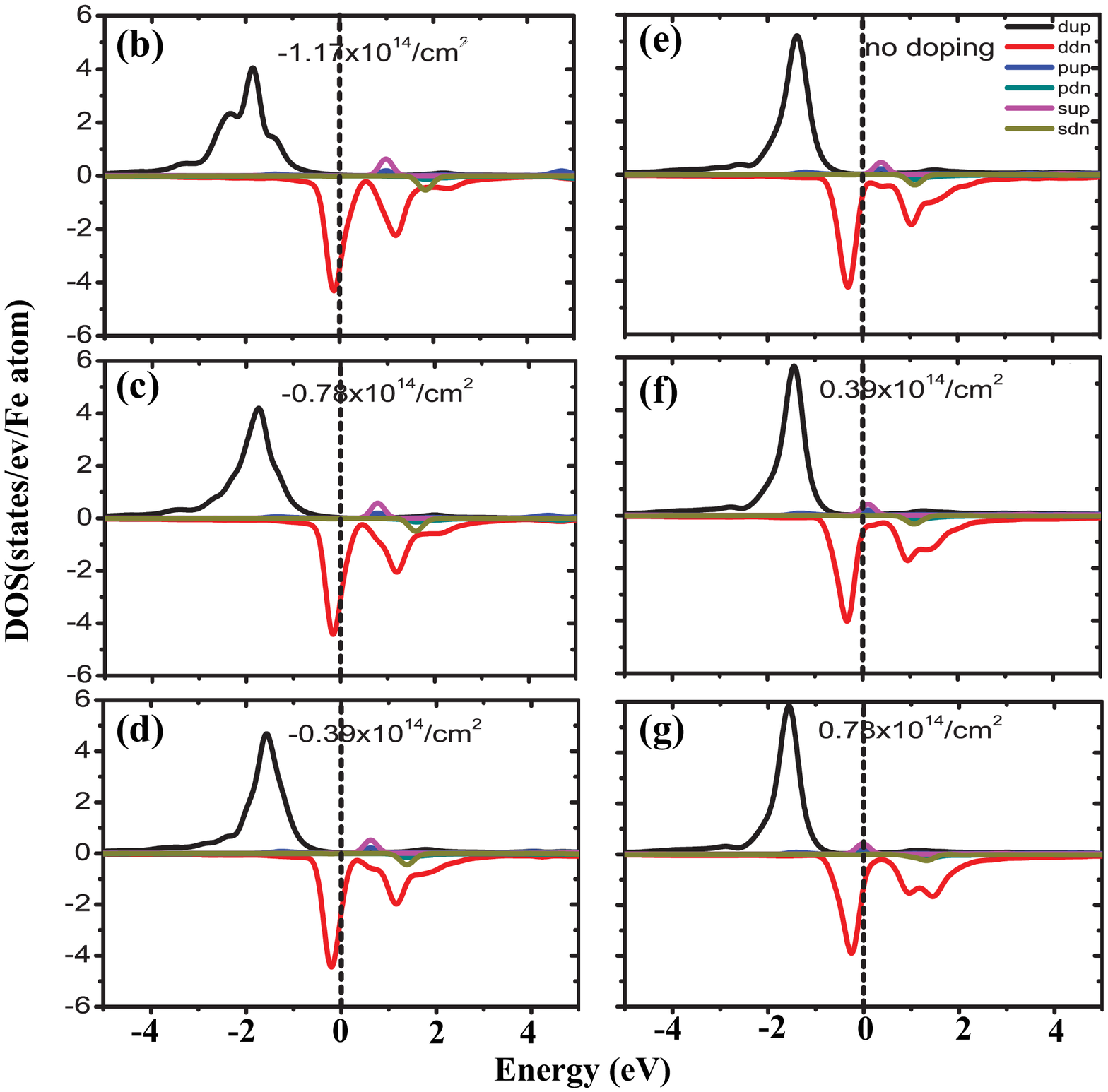}
\caption{(color online) (a) Local magnetic moment of Fe adatom of
H-site versus charge doping concentration; (b-g) Fe adatom partial
density of states with projection to s, p, d orbitals for both
spin-up and spin-down components.}
\end{figure}

The Fe-adatom diffusion barrier is shown in In Fig. 1(c) as a
function of charge doping concentration. Without charge doping, the
diffusion barrier is 0.48~eV \cite{barrier}, in good agreement with
previous report. With hole doping the diffusion barrier can be
increased to 0.55~eV but with electron doping diffusion barrier can
be decreased to 0.28~eV. This trend can be again understood from the
charge doping effect on the adsorption energy of Fe at H-site and
B-site. The diffusion barrier is the adsorption energy difference
between Fe at B-site and Fe at H-site, we thus express diffusion
barrier E$_d$:
\begin{eqnarray}
E_{d}(q)=E_r^B(q)-E_r^H(q)-(q_B-q_H)\phi
\end{eqnarray}
The first order variation of $E_{d}$ in charge doping will then be:
\begin{align}
\Delta E_{d}(q)=&\Delta(E_r^B(q)-E_r^H(q))-(q_B-q_H)\Delta\phi
\\ \nonumber &-(\Delta q_B-\Delta q_H)\phi
\end{align}
The second term indicates that a direct tuning of work function
$\phi$ will lead to a variation of diffusion barrier depending on
the sign of work function change and the magnitude. Because work
function is increased with hole doping, this term gives rise to an
increase of diffusion barrier. On the other hand, because work
function is decreased with electron doping, this term gives rise to
an decrease of diffusion barrier. This predication is consist with
the trend of diffusion barrier in Fig. 1(c) calculated from DFT. We
thus believe the work function tuning should be the dominant role in
varying the Fe-adatom diffusion barrier.

For no-charge doping graphene+Fe adatom, previous work \cite{Femag1,
Femag2} has shown that because of the hybridization between Fe 3d
states and graphene p states, the Fe 4s states are shifted to higher
energy relative to Fe 3d states upon adsorption and originally two
occupying two 4s electrons are transferred mainly to Fe 3d states,
resulting in the Fe local magnetic moment reduction from $4~\mu_B$
to about $2~\mu_B$. Such a situation is expected to be further
modified upon charge doping, which may change the Fe adatom orbital
occupation. In Fig. 3(a) we show the Fe adatom local magnetic moment
versus the charge doping concentration. Hole doping significantly
increases the magnetic moment from $2.05~\mu_B$ to $2.73~\mu_B$, and
electron doping modestly increases the magnetic moment to
$2.32~\mu_B$. In Fig. 3(b) the density of states under different
charge doping concentration are plotted, from which we can see the
different Fe adatom orbital occupation which leads to the Fe
magnetic moment variation with charge doping. Starting from
zero-doping to increasing hole doping, the occupation of spin-down
component of Fe d-orbital keeps decreasing and the occupation of
spin-up component is unchanged. This leads to the further imbalance
between spin-up and spin-down states and leads to increased Fe
magnetic moment. With increasing electron doping, the slight
decrease in Fe spin-down d-orbital occupation and increase in Fe
spin-up s-orbital result in the slow increase of Fe magnetic moment.

\begin{figure}
\includegraphics[clip,scale=0.50]{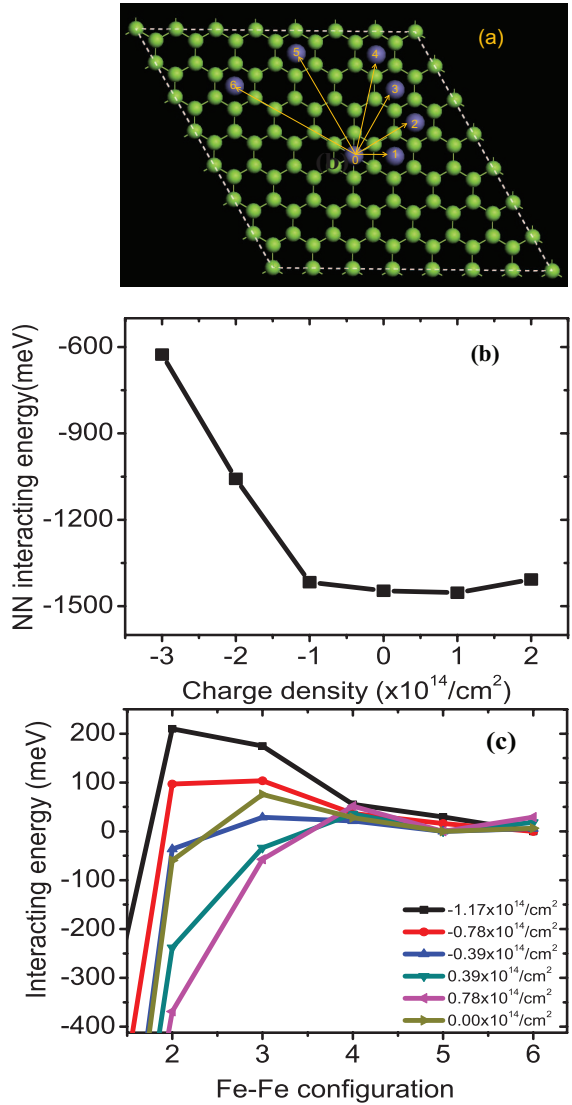}
\caption{(color online) (a) Interaction energy between two
nearest-neighboring Fe adatoms; (b) Interaction energy as a function
of separation beyond nearest-neighbor.}
\end{figure}

Next we calculated Fe adatom-adatom interaction energy as a function
of the separation under different charge doping. Six configurations
are considered as shown in Fig. 4(a). For clarity we separated
nearest-neighbor (NN) adatom-adatom interaction (configuration 1)
which represents the direct chemical bonding from the beyond NN
adatom-adatom interaction. They are shown in Fig. 4(b) and Fig.
4(c), respectively. The NN interaction energy is only changed very
little under the charge doping concentration from
$-0.39\times10^{14}/cm^2$ to $0.78\times10^{14}/cm^2$. However, one
observes that further hole doping decreases the NN interaction
energy rapidly and the the NN interaction energy is reduced a lot to
0.60~eV under hole doping of $-1.17\times10^{14}/cm^2$ with respect
to 1.45~eV for no-doping situation. Recalling the Bader charge in
Fig. 1(b), because the Bader charge keeps decreasing from hole
doping to electron doping, we may attribute this reduction of NN
interaction energy to the significantly increased repulsive Coulomb
interaction under large hole doping and only a lot weaker
dipole-dipole repulsive interaction under the doping concentration
from $-0.39\times10^{14}/cm^2$ to $0.78\times10^{14}/cm^2$. In Fig.
4(c) within the doping concentration considered in the work, the
adatom-adatom distance at which they display repulsive interaction
persistly exists. Apparently, the repulsive peak is pushed gradually
towards to be at next NN distance and increased from electron doping
to hole doping.

\begin{figure}
\includegraphics[clip,scale=0.30]{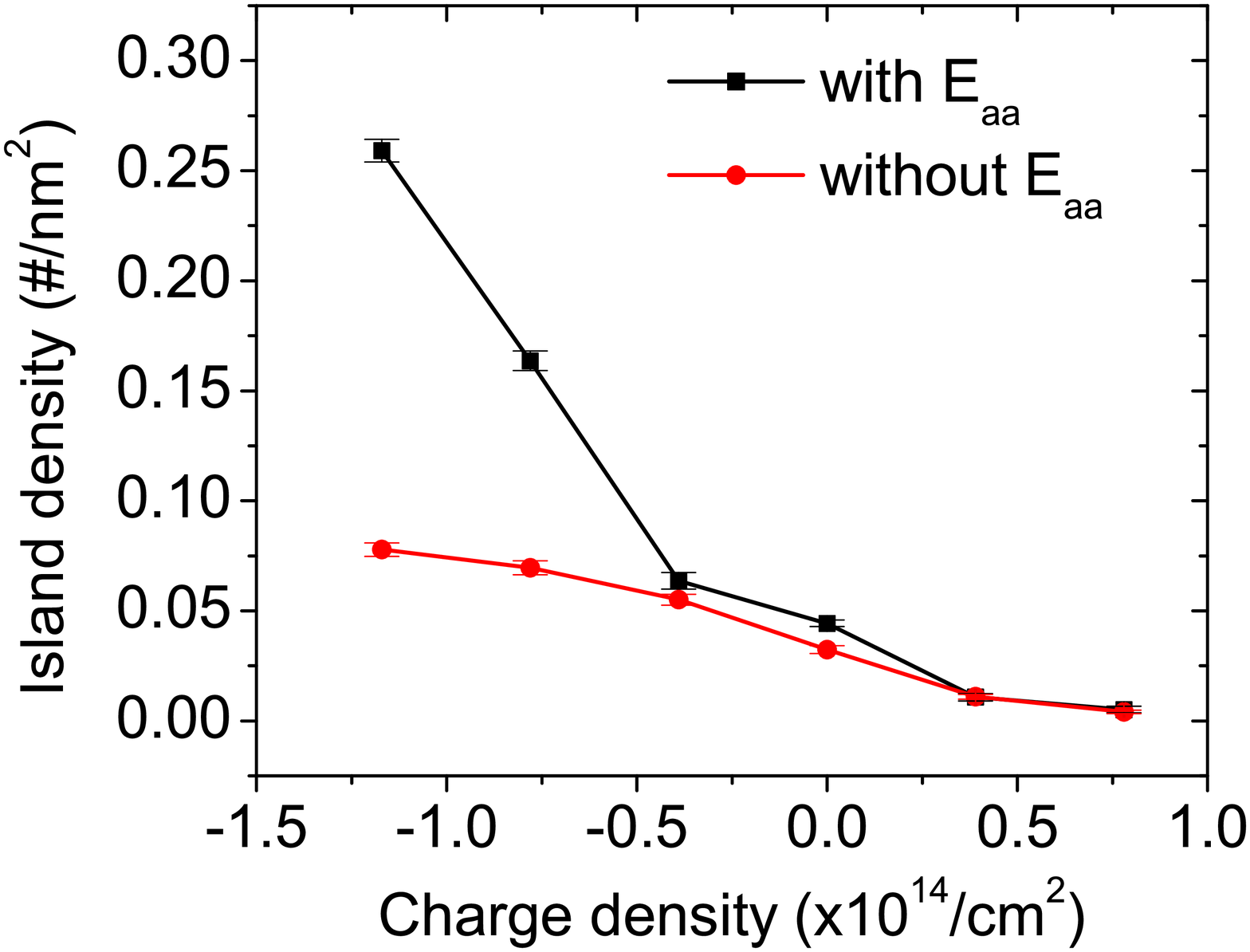}
\caption{(color online) KMC simulated island density as a function
of charge doping concentration for both cases of with adatom-adatom
interaction (with $E_{ad}$) and without adatom-adatom interaction
(without $E_{ad}$).}
\end{figure}

The charge doping effects on the Fe-adatom diffusion barrier and
adatom-adatom interaction is expected to be reflected in the Fe
nucleation island density of the initial film growth. We next
simulated the Fe island density as a function of charge doping
concentration using kMC simulation method proposed in
Ref.\cite{kmc}. The simulation cell is $200x200$ graphene supercell.
The diffusion barrier and adatom-adatom interactions from DFT
calculations above are used as input parameters. The hopping rate
with Arrhenius form of $\nu=\nu_0\exp(-E_d/k_BT)$ and position
dependent diffusion barrier approximation of
$E_d=E_d^0+0.5(E_j-E_i)$ were used. $\nu_0$ is chosen to have
constant value of $10^{12}/s$, T is 300~K, E$_i$ and E$_j$ are the
interaction energies before and after hopping, respectively. For
simplicity irreversible nucleation (no desorption ), critical island
size of 1 and no edge diffusion are assumed \cite{island1, island2}.
The deposition rate is 0.01~ML/s and amount of deposition is
0.05~ML. In Fig. 5 we show the island density for both situations
with and without including Fe adatom-adatom interaction. From the
curve without adatom-adatom interaction, the island density can be
decreased to $~8$ times smaller in electron doping and $~3$ times
larger in hole doping than in zero-doping. Including Fe
adatom-adatom interaction, it's most evident for the hole doing
larger than $-0.39\times10^{14}$, the island density is
significantly increased up to $~6$ times larger than in zero-doping.
For the remaining doping regime, the island density is very close to
that without inter-adatom interaction. It indicates that the
combined effect of the diffusion barrier and inter-adatom
interaction on the island density only takes place in large hole
doping and the diffusion barrier tuning dominates the change of
island density in the rest of the charge doping regime.

To conclude, we have investigated the effect of the charge doping of
graphene substrate on Fe nucleation island density, which increases
under hole-doping and decreases under electron-doping. The
underlying mechanism is from the charge-tuning of Fe-adatom
diffusion barrier, which is gradually increased by hole doping but
is rapidly decreased by electron doping, and Fe inter-adatom
repulsive interaction, which is increased significantly by large
hole doping. Additionally Fe local magnetic moment can be tuned
significantly with charge doping. The combined effects provide large
range of tuning of magnetic island density and tailoring the growth
morphology of magnetic metal nanostructure for spintronics
applications via surface charge doping.

This work was supported by NSF MRSEC (Grant No. DMR-1121252) and
DOE-BES (Grant No. DE-FG02-04ER46148). We thank M. C. Tringides and
Y. Han and for fruitful discussions. We acknowledge the CHPC at
University of Utah and NERSC for providing the computing resources.


\begin{references}

\bibitem{carrier1}
Aaron Bostwick \emph{et al}, Nat. Phys. {\bf 3}, 36 (2007).

\bibitem{carrier2}
S. Y. Zhou \emph{et al}, Phys. Rev. Lett. {\bf 101}, 086402 (2008).

\bibitem{mag1}
T. Eelbo \emph{et al}, Phys. Rev. Lett. {\bf 110}, 136804 (2013).

\bibitem{mag2}
A. V. Krasheninnikov \emph{et al}, Phys. Rev. Lett. {\bf 102},
126807 (2009).

\bibitem{catalyst1}
Shaojun Guo \emph{et al}, Angew. Chem. Int. Ed. {\bf 51}, 11770
(2012).

\bibitem{catalyst2}
Kun Han \emph{et al}, Appl. Phys. Lett. {\bf 104}, 053101 (2014).

\bibitem{super1}
S. -L. Yang \emph{et al}, Nat. Commun. {\bf 5}, 3493 (2014).

\bibitem{super2}
Gianni Profeta \emph{et al}, Nat. Phys. 8, {\bf 131} (2012).

\bibitem{subc1}
C. Riedl \emph{et al}, Phys. Rev. Lett. {\bf 103}, 246804 (2009).

\bibitem{Efie}
Kin Fai Mak \emph{et al}, Phys. Rev. Lett. {\bf 112}, 207401 (2014).

\bibitem{subdop}
Nisha Mammen \emph{et al}, J. Am. Chem. Soc. {\bf 133}, 2801 (2011).

\bibitem{cbond}
J. O. Sofo \emph{et al}, Phys. Rev. B {\bf 83}, 081411(R) (2011).

\bibitem{adad}
Dmitry Solenov \emph{et al}, Phys. Rev. Lett. {\bf 111}, , 115502
(2013).

\bibitem{oxideT1}
Livia Giordano \emph{et al}, Phys. Rev. Lett. {\bf 101}, 026102
(2008).

\bibitem{corr1}
Kristen A. Fichthorn \emph{et al}, Phys. Rev. Lett. {\bf 84}, 5371
(2000).

\bibitem{corr2}
Jascha Repp \emph{et al}, Phys. Rev. Lett. {\bf 85}, 2981 (2000).

\bibitem{corr3}
A. Bogicevic \emph{et al}, Phys. Rev. Lett. {\bf 85}, 1910 (2000).

\bibitem{corr4}
Fabien Silly \emph{et al}, Phys. Rev. Lett. {\bf 92}, 016101 (2004).

\bibitem{Fehigh}
S. M. Binz \emph{et al}, Phys. Rev. Lett. {\bf 109}, 026103 (2012).


\bibitem{FeFerepul}
Xiaojie Liu \emph{et al}, Phys. Rev. B {\bf 84}, 235446 (2011).

\bibitem{PAW}
G. Kresse and D. Joubert, Phys. Rev. B {\bf 59}, 1758 (1999).

\bibitem{GGA}
J. P. Perdew, J. A. Chevary, S. H. Vosko, K. A. Jackson, M. R.
Pederson, D. J. Singh and C. Fiolhairs, Phys. Rev. B {\bf 46}, 6671
(1992).

\bibitem{vasp}
G. Kresse and J. Furthmuller, Phys. Rev. B {\bf 54}, 11169 (1996).


\bibitem{NEB}
G. Mills and H. Jonsson, Phys. Rev. Lett. {\bf 72}, 1124 (1994).

\bibitem{barrier}
Oleg V. Yazyev \emph{et al}, Phys. Rev. B {\bf 82}, 045407 (2010).

\bibitem{Femag1}
Chao Cao \emph{et al}, Phys. Rev. B {\bf 81}, 205424 (2010).
\bibitem{Femag2}
Kevin T. Chan \emph{et al}, Phys. Rev. B {\bf 77}, 235430 (2008).

\bibitem{kmc}
K. A. Fichthorn and W. H. Weinberg, J. Chem. Phys. 95, 1090 (1991).

\bibitem{island1}
M. C. Bartelt and J. W. Evans, Phys. Rev. B {\bf 46}, 12675 (1992).

\bibitem{island2}
H. Brune, Surf. Sci. Rep. {\bf 31}, 121 (1998).

\end{references}
\end{document}